\documentclass[12pt]{article}
\usepackage{amsmath}
\usepackage{amsfonts}
\usepackage{amssymb}
\usepackage{mltex}
\usepackage{graphicx}
\newtheorem{lemma}{Lemma}[section]
\newtheorem{theorem}[lemma]{Theorem}
\newtheorem{proposition}[lemma]{Proposition}
\newtheorem{corollary}[lemma]{Corollary}
\newtheorem{remark}[lemma]{Remark}

\newtheorem{definition}[lemma]{Definition}

\def\sq{\hbox {\rlap{$\sqcap$}$\sqcup$}}
\def\sq{\hbox {\rlap{$\sqcap$}$\sqcup$}}

\def\1{{\rm 1\mskip-4.5mu l} }
\def\lsim{\raise0.3ex\hbox{$<$\kern-0.75em\raise-1.1ex\hbox{$\sim$}}}
\def\gsim{\raise0.3ex\hbox{$>$\kern-0.75em\raise-1.1ex\hbox{$\sim$}}}

\def\beq{\begin{equation}}   \def\edq{\end{equation}}
\def\bea{\begin{eqnarray}}  \def\eea{\end{eqnarray}}

\newcommand\mysection{\setcounter{equation}{0}\section}
\renewcommand{\theequation}{\thesection.\arabic{equation}}
\newcounter{hran} \renewcommand{\thehran}{\thesection.\arabic{hran}}

\def\bmini{\setcounter{hran}{\value{equation}}
    \refstepcounter{hran}\setcounter{equation}{0}
    \renewcommand{\theequation}{\thehran\alph{equation}}\begin{eqnarray}}

\def\bminiG#1{\setcounter{hran}{\value{equation}}
\refstepcounter{hran}\setcounter{equation}{-1}
\renewcommand{\theequation}{\thehran\alph{equation}}
\refstepcounter{equation}\label{#1}\begin{eqnarray}}

\def\emini{\end{eqnarray}\relax\setcounter{equation}{\value{hran}}\renewcommand{
\theequation}
{\thesection.\arabic{equation}}}

\begin{document}

\title {The Mutually Unbiased Bases Revisited}
%\end {center}
\author{\it {\bf Monique Combescure} \\
\it IPNL, B\^atiment Paul Dirac \\
\it 4 rue Enrico Fermi,Universit\'e Lyon-1 \\
\it  F.69622 VILLEURBANNE Cedex, France\\
\it email monique.combescure@ipnl.in2p3.fr}
\vskip 1 truecm
\date{}
\maketitle

\begin{abstract}
The study of Mutually Unbiased Bases continues to be developed vigorously, and presents
several challenges in the Quantum Information Theory.\\
Two orthonormal bases in $\mathbb C^d,\ B\ \mbox{and}\ B'$ are said mutually unbiased
if $\forall b\in B,\ b'\in B'$ the scalar product $b\cdot b'$ has modulus $d^{-1/2}$.
In particular this property has been introduced in order to allow an optimization of the
 measurement-driven quantum evolution process of any state $\psi \in
\mathbb C^d$ when measured in the mutually unbiased bases $B_{j}\ \mbox{of}\ 
\mathbb C^d$.\\
At present it is an open problem to find the maximal umber of mutually Unbiased Bases when
$d$ is not a power of a prime number.\\

\noindent
In this article, we revisit the problem of finding Mutually Unbiased Bases (MUB's) in any
dimension $d$. The method is very elementary, using the simple unitary matrices introduced
by Schwinger in 1960, together with their diagonalizations. The Vandermonde matrix
based on the $d$-th roots of unity plays a major role. \\
This allows us to show the existence of a set of 3 MUB's in any dimension, to give conditions for existence
of more than 3 MUB's for $d$ even or odd number, and to recover the known result of
existence of $d+1$ MUB's for $d$ a prime number. Furthermore the construction of these
MUB's is very explicit. \\
As a by-product, we recover results about Gauss Sums, known in number theory, but which have 
apparently not been previously derived from MUB properties.

\end{abstract}
\newpage

\mysection{INTRODUCTION}
Two orthonormal bases $B$ and $B'$ in $\mathbb C^d$ are called mutually unbiased if
$\vert b\cdot b'\vert = d^{-1/2}, \ \forall b \in B,\ b' \in B'$, where $v\cdot v'$
denotes the scalar product in $\mathbb C^d$. This notion first appeared in the literature in
\cite{schwinger} in 1960, although the term ``Mutually Unbiased Bases'' (MUB) appeared
later. It has attracted recently a great interest in the physics as well as mathematics literature,
in conjunction with questions of Quantum Information, Quantum Cryptography, and
Quantum Entanglement (see \cite{cha1}, \cite{cha2}, \cite{iva}, \cite{klaro}, 
\cite{plaro}, \cite{wofi}, \cite{wootters}, and references therein contained).
 Note in particular that this property has been developed in order to allow an optimization of 
 the measurement-driven quantum evolution process of any state $\psi \in
\mathbb C^d$ when measured in the mutually unbiased bases $B_{j}\ \mbox{of}\ 
\mathbb C^d$ \cite{roa}, \cite{wofi}.\\
If we denote by $N(d)$ the maximum cardinality of a set of MUB in $\mathbb C^d$, it
has been established that 
$$N(d)\le d+1$$
with equality for $d$ being a power of a {\bf prime number} (see \cite{wofi}, \cite{iva},
\cite{band}, \cite{klaro} and references herein contained).\\
In a number of previous works (see for example \cite{ar}, \cite{band}, \cite{cha1},
\cite{cha2}, \cite{kipla}, \cite{klaro}, \cite{plaro}, \cite{wootters}),
 it has been recognized that the construction of MUB's has to do
with rather sophisticated arithmetical notions such as Weil sums over finite fields, Gauss Sums
and Galois rings.\\
In this paper, we revisit these known results from an elementary point of view based on 
a simple set of $d \times d$ of unitary matrices. 
In \cite{kipla}, a recipe for an explicit constuction of the set of all MUB's for $d$ a power of a
 prime has been provided, using the angular momentum bases.
Strongly inspired by the recent work of
Kibler and Planat \cite{kipla}, we reintroduce the matrices constructed by Schwinger, which allows
us a construction of MUB's in different cases:\\
- d any integer\\
- d an odd integer\\
- d a prime number.\\
The three building block of unitary matrices that allow to perform our construction are, if
$q :=\exp(\frac{2i\pi}{d})$ the following:
$$U:= {\rm diag}(1,q,q^2,...,q^j,...,q^{d-1})$$

$$V:= \left(
\begin{array}{cccccc}
0&1&0&.&.&0\\
0&0&1&.&.&0\\
.&.&.&.&.&.\\
.&.&.&.&.&.\\
0&0&0&.&.&1\\
1&0&0&.&.&0
\end{array}\right)$$
and for $d$ an odd number
$$D:= {\rm diag}(1,q, q^3,...q^{\frac{j(j+1)}{2}}, ...1)$$
The result is that the diagonalization of the $d$ matrices $V_{k}:= VU^k, \ k \in\left\{
0,1,...,d-1\right\}$
(also studied in \cite{kipla}) provides us with a set of unitary matrices $P_{k}$ which have
certain ``unbiasement'' properties, according to the various cases listed above. A similar
idea is also developped in \cite{band} where the matrices $U$ and $V$ are called
``generalized Pauli matrices on $d$-state quantum systems''.\\
As a by-product, we recover certain properties of Gauss Sums, which to our knowledge
has not been deduced from previous studies on MUB (see however the recent
work \cite{kipla} where a similar but different sum rule appears for $d$ a prime number).

$$\left\vert \sum_{j=0}^{d-1}q^{\frac{kj(j+1)}{2}}\right\vert =
 \sqrt d, \ \mbox{if}\ d\  \mbox{is odd}\ \mbox{and}\ \forall 
  k\ \mbox{coprime with}\ d$$
 This property can be found in the number theory literature (\cite{berndt}).

\mysection{THE SCHWINGER MATRICES}

\subsection{GENERAL DEFINITIONS AND PROPERTIES}

In \cite{schwinger},  two basic unitary $d\times d$ matrices $U,\ V$ are introduced.
Let 
\beq
\label{q}
q:= \exp\left(\frac{2i\pi}{d}\right)
\edq
They are of the following form:
\beq
\label{U}
U:={\rm Diag}(1,q,q^2,..., q^{d-1})
\edq
\beq
\label{V}
V:= \left(
\begin{array}{cccccc}
0&1&0&.&.&0\\
0&0&1&.&.&0\\
.&.&.&.&.&.\\
.&.&.&.&.&.\\
0&0&0&.&.&1\\
1&0&0&.&.&0
\end{array}\right)
\edq

\begin{lemma}
(i) $U,\ V$ obey the ``q-commutation rule'':
\beq
\label{qcom}
VU=qUV
\edq
(ii) The Vandermonde matrix $P_{0}$ whose matrix elements for $j,k \in \left\{0,1,...,
d-1\right\}$ are defined by
\beq
\label{VDM}
(P_{0})_{j,k}:= d^{-1/2}q^{jk}
\edq
is such that
\beq
\label{diagV}
V=P_{0}UP_{0}^*
\edq
\end{lemma}

\begin{definition}
For any $k\in \left\{0,1,...,d-1\right\}$ we define:
\beq
\label{Vk}
V_{k}:=VU^k = \left(
\begin{array}{cccccc}
0&q^k&0&.&.&0\\
0&0&q^{2k}&.&.&0\\
.&.&.&.&.&.\\
.&.&.&.&.&.\\
0&0&0&.&.&q^{k(d-1)}\\
1&0&0&.&.&0
\end{array}\right)
\edq
\end{definition}

\begin{remark}
The matrices $V_{k}$ have been first introduced in the study of MUB by Kibler-Planat
\cite{kipla}. 
\end{remark}

\begin{definition}
(i) We say that a $d \times d$ unitary matrix $A$ is ``unbiased'' is all its matrix elements $A_{j,k}$
sasisfy
\beq
\label{biased}
\vert A_{j,k}\vert = d^{-1/2}, \ \forall j,k \in \left\{0,1,..., d-1\right\}
\edq
(ii) We say that two $d \times d$ unitary matrices $A,\ B$ are ``mutually unbiased'' if the
 matrix $A^*B$ is unbiased.
\end{definition}

Thus finding a MUB in dimension $d$ amounts to exhibit a set that we call a MUM, of the 
following form:
\beq
\label{MUM}
\left\{ \1_{d}, P_{0}, P_{1},..., P_{m}\right\}
\edq
(where $\1_{d}$ denotes the identity $d\times d$ matrix) such that $P_{j}, \ j \in 
\left\{0,1,...,m\right\}$ are ``unbiased'', and $P_{j},\ P_{k},\ j,k \in \left\{
0,1,...,m\right\},\ j\ne k$ are ``mutually unbiased''.

\begin{proposition}
(i) Let , for any $k \in \left\{0,1,..., d-1\right\}$, $P_{k}$ be a unitary $d \times d$ matrix, and
$D_{k}$ be the unitary diagonal matrix such that
\beq
\label{diagVk}
V_{k}= P_{k}D_{k}P_{k}^*
\edq
Then all matrices $P_{k}$ are ``unbiased matrices''.\\
(ii) Furthermore $D_{0}\equiv U$.
\end{proposition}

\begin{lemma}
For any $k \in \left\{0,1,...,d-1\right\}$ one has
\beq
\label{diagVk}
U^k P_{0}= P_{0}(V^*)^k
\edq
\end{lemma}

Proof: It is known \cite{berndt} (and easy to check) that
$P_{0}^2=W$ where $W\equiv W^*$ is the permutation matrix
\beq
\label{perm}
W:= \left(
\begin{array}{ccccccc}
1&0&0&0&.&.&0\\
0&0&0&0&.&.&1\\
.&.&.&.&.&.&.\\
.&.&.&.&.&.&.\\
0&0&1&0&.&.&0\\
0&1&0&0&.&.&0
\end{array}\right)
\edq
We want to prove that:
$$V^*=P_{0}^*UP_{0}$$
But using Lemma 2.2, this is equivalent to:
$$V^*= P_{0}^{*2}VP_{0}^2\equiv WVW$$
which follows immediately from the property of the selfadjoint matrix $W$ that:
$$WV^*=VW$$
Thus we have proven (\ref{diagVk}) for $k=1$. The general statement follows by induction
since:
\beq
\label{100}
U^kP_{0}= U U^{k-1}P_{0}= UP_{0}(V^*)^{k-1}= P_{0}(P_{0}^*UP_{0})(V^*)^{k-1}
=P_{0}(V^*)^k
\edq

\begin{proposition}
For {\bf any dimension $d\ge 2$},  if $P_{1}$ be a unitary $d \times d$ matrix such that
$$V_{1}= P_{1}D_{1}P_{1}^*$$
then the matrices $P_{1},\ P_{0}$ are mutually unbiased $d \times d $ matrices. 
\end{proposition}

Proof: One has, using Lemma 2.1 (ii) and Lemma 2.6 for $k=1$ that:
$$P_{0}^*V_{1}P_{0} =P_{0}^* VUP_{0}= P_{0}^*VP_{0}V^*= UV^*$$
Thus
$$P_{0}^*P_{1}D_{1}P_{1}^*P_{0}= UV^*$$
which means that all column vectors of $P_{0}^*P_{1}$ are eigenstates of $UV^*$ with
eigenvalues being the diagonal elements of $D_{1}$ which are all of modulus 1. Since
any eigenstate $v:= (v_{0},v_{1},..., v_{d-1})$ of the matrix $UV^*$ satisfy $\vert v_{j}
\vert =\vert v_{j}\vert, \ \forall j,k \in \left\{0,1,..., d-1\right\}$ and $P_{0}^*P_{1}$
is unitary, this implies the result.\\
\sq\\

\begin{corollary}
For any integer $d\ge 2$, there is {\bf at least three MUB} given by the bases defined by $\1_{d}, P_{0}
, P_{1}$.
\end{corollary}

The existence of at least 3 MUB's in any dimension is proven in \cite{klaro}.

\subsection{THE EVEN CASE}
Let $d$ be {\bf even}. Then the determinant of both $U,\ V$ equals $\pm 1$ depending on
whether $d=0\ \mbox{or}\ 2\ (\mbox{mod}\ 4)$. Namely
$$\det U = q^{\frac{d(d-1)}{2}}$$
and $d(d-1)/2$ is half integer if $d=2\ (\mbox{mod}\ 4)$, and integer if $d=0\ (\mbox{mod}\ 4)$.
\\
In both cases the matrix $V_{1}=VU$ has thus determinant +1, which means that it is 
unitarily equivalent to $\omega U$, where 
$$\omega:= \exp\left(\frac{i\pi}{d}\right)$$
The eigenstate $v^{(1)}:= (1,a_{1}, a_{2},..., a_{d-1})$ of $V_{1}$ with eigenvalue 
$\omega$ is such that $a_{1}= \omega^{-1}= a_{d-1}$, and obeys the recurrence relation
$$a_{k}= \omega^{1-2k}a_{k-1}$$
Thus solving he recurrence relation we have:
$$a_{k}= \omega^{\sum_{j=0}^k (1-2j)}= \omega^{k-k(k+1)}= \omega^{-k^2}$$
More generally the eigenstate $v^{(j)}:= (1, b_{1}, ..., b_{k}, ..., b_{d-1})$ of $V_{1}$
with eigenvalue $\omega^{2j+1}$ is such that
$$b_{k}= \omega^{2jk-k^2}\equiv q^{jk}\omega^{-k^2}$$
This implies:

\begin{proposition}
(i) The matrix $P_{1}$ defined by:
$$P_{1}= D' P_{0}$$
with
$$D':= {\rm diag}(1,\omega^{-1},..., \omega^{-k^2},..., \omega^{-1})$$
diagonalizes $V_{1}$, namely $D_{1}\equiv \omega U$:
$$V_{1}= \omega P_{1}UP_{1}^*$$
(ii) The property already shown that $P_{0},\ P_{1}$ are mutually unbiased reflects itself
in the identity
$$\vert {\rm Tr}D'\vert=\left\vert \sum_{k=0}^{d-1}\omega^{k^2}\right\vert = \sqrt d$$
\end{proposition}

The proof of (i) is obvious.
Furthermore (ii) results from a known property in number theory
\cite{berndt}, that if $d$ is even, then
$$\sum_{k=0}^{d-1}\exp\left(k^2\frac{i\pi}{d}\right)= \sqrt d \exp\left(
\frac{i\pi}{4}\right)$$
\sq\\

\noindent
For $d$ even but not {\bf not a power of 2}, it is not known what is the maximum
number of MUB's. For example for $d=6$ there is a conjecture that $N(6)=3$ (see Section 6
where an explicit set of 3 MUB's is constructed). For $d=0\ (\mbox{mod 4})$,
it is known that the ``tensor-product method'' provides sets of more than 3 MUB's (see \cite{klaro}).
In Section 7, we make explicit this construction of 4 (resp 5)  MUB's in the case $d=12$
(resp $d=20$).

\subsection{THE ODD CASE}

\begin{definition}
Let us define $F_{d}:= \mathbb Z / d\mathbb Z$ which is the finite field of residues of
$n, \ (\mbox{mod}\ d)$.
\end{definition}

\begin{theorem}
Let $d \in \mathbb N$ be an {\bf odd} number. Define the unitary diagonal matrix $D$ as
\beq
\label{D}
D:= {\rm diag}(1,q, q^3,...q^{\frac{j(j+1)}{2}}, ...1)
\edq
Then we have:\\
(i) The matrices $V_{k}, \ k \in F_{d}$ are all unitarily equivalent to $U$.\\
(ii) Let $P_{k}:= D^{-k}P_{0}$; then, for all $k \in F_{d}$ one has:
$$P_{k}^*V_{k}P_{k}= U$$ 
In other words if $P_{0}= (v_{0}, v_{1},...,v_{d-1})$, then
$$P_{k}^*= (v_{0}, q^kv_{d-1}, ..., q^{kj(j+1)/2}v_{d-j},..., v_{1})$$
(iii) $\forall k \in F_{d}, \ \mbox{such that}\ d,\ k$ are co-prime, one has
\beq
\label{tr}
\vert {\rm Tr}D^{k}\vert = \sqrt d
\edq
\end{theorem}

Proof: (i) is a consequence of (ii). Let us prove (ii):\\
It is enough to check that
$$U= P_{0}^*D^kVU^kD^{-k}P_{0}$$
But $U^k$ and $D^{-k}$ being diagonal commute, so that we are left with
$$D^kVD^{-k}U^k =P_{0}UP_{0}^*$$
this in turn is equivalent to
$$D^kVD^{-k}= VU^{-k}\equiv V_{-k}$$
or to the equation
$$D^kV=V_{-k}D^k$$
which follows easily from the fact that they are unitary matrices with only non-vanishing 
elements $a_{0,d-1}=1$ and
$$a_{j,j+1}= \left(q^{\frac{k(k+1)}{2}}\right)^k,\ \forall j\in \left\{0,1,...,d-1
\right\}$$
Now let us prove (iii). We need the following proposition:

\begin{proposition}
Let $k \in F_{d},\ \mbox{such that}\ k,\ d $ are co-prime. Then the matrix
$P_{0}^*P_{k}$ is unbiased.
\end{proposition}

Proof: It follows from equ. (\ref{100}) that
 $$V_{k}P_{0}\equiv VU^k P_{0}= VP_{0}(V^*)^k$$
 and thus
 \beq
 \label{101}
 P_{0}^*P_{k}UP_{k}^*P_{0}= U(V^*)^k
 \edq
 (since by definition $V_{k}= P_{k}UP_{k}^*$)\\
But:

\begin{lemma}
 If $d,\ k \in F_{d}$ are co-prime, the matrix $(V^*)^k$ is a permutation matrix with cycle of length
 $d$, and thus all eigenstates of $U(V^*)^k$ have coordinates of equal modulus, namely
 $d^{-1/2}$,
\end{lemma}

Proof: This is standard. For any $d,\ k \in F_{d}$ co-prime, there exists a cyclic permutation 
 $\sigma_{k}$ (that means a permutation with cycle of length $d$)
of $F_{d}$ such that for any $v \in \mathbb C^k$, the element $w \in \mathbb C^k$
defined by:
$$(V^*)^kv \equiv w$$
is such that
$$w_{j} = v_{\sigma_{k}(j)}, \ \forall j \in F_{d}$$
\sq\\

\begin{remark}
 The idea that the eigevectors of $V_{k}$ are ``cyclically shifted'' modulo a phase if $d$
 is a prime number has already been put forward in \cite{band}.\\
\end{remark}

End of Pooof of Proposition 2.12:\\
 Let us denote by $v^{(k)}$ the successive column vectors of $P_{0}^*P_{k}$. Then
$$P_{0}^*P_{k}U= (q^0v^{(0)}, qv^{(1)},..., q^jv^{(j)},...,q^{d-1}v^{(d-1)})$$
This means that $v^{(j)}$ is eigenvector of the matrix $U(V^*)^k$ with eigenvalue $q^j$.
Therefore we have that 
 $\vert v^{(j)}_{l}\vert = \vert v^{(j)}_{0}\vert,\ \forall l \in \left\{0,1,...,
d-1\right\}$, as a consequence of Lemma 2.13 above. Since $\Vert v \Vert =1$, this
 implies $\vert v^{(j)}_{k}\vert = d^{-1/2}$.
 It follows that for all primes $k\ \in F_{d}$ that are relatively prime to $d$, one has that
 $P_{k}^*P_{0}$ is an unbiased matrix.\\
 \sq\\
 
 Proof of Theorem 2.11 (iii):\\
 Let $d,\ k \in F_{d}$ be co-prime. Let us call $v_{k}$ the normalized
eigenvector of $V_{k}$ with eigenvalue 1.
We obviously have
$$(v_{k})_{j}= \frac{1}{\sqrt d}\left(q^{\frac{j(j+1)}{2}}\right)^k$$
Now using that $P_{0}^*P_{k}$ is unbiased we have
$\vert v_{0}\cdot v_{k}\vert= d^{-1/2}$ and
$$v_{0}\cdot v_{k}\equiv d^{-1}\sum_{j=0}^{d-1}(q^{\frac{j(j+1)}{2}})^k
\equiv d^{-1}{\rm Tr}(D^k)$$
which yields the result.
\sq\\

\begin{corollary}
Let d be an {\bf odd number}. Then for any $k \in F_{d}$ co-prime with $d$, we have:
$$\left\vert \sum_{j=0}^{d-1}q^{\frac{kj(j+1)}{2}}\right\vert =
 \sqrt d$$
\end{corollary}

\begin{remark}
Corollary 2.12 is strongly related to the property of Gauss Sums. In \cite{berndt}, the
following result is established: define, for $a,b,d \in \mathbb Z,\ \mbox{with}\ ad+b\ 
\mbox{even, and}\ ad\ne 0$
$$S(a,b,d):= \sum_{n=0}^{d-1}\exp\left(\frac{i\pi(an^2+bn)}{d}\right)$$
Then the following ``reciprocity theorem for quadratic Gauss sums'' yields that:
\beq
\label{recipr}
S(a,b,d)= \left\vert \frac{d}{a}\right\vert\exp\left(\frac{i\pi}{4}
(\mbox{sgn}(ad)-b^2/ad)\right)S(-d,-b,a)
\edq
Applying it with $d$ {\bf odd} and $a=b=1$, we have
$$S(1,1,d)= \sqrt d \exp\left(\frac{i\pi}{4}(1-\frac{1}{d})\right)$$
since $S(-d,1,1)=1$.\\
Thus arithmetics gives not only the modulus of ${\rm Tr}D$ which equals $\sqrt d$
but also the phase. A similar result holds for ${\rm Tr}D^k$ provided $d,\ k\in F_{d}$ 
are co-prime.
\end{remark}

If $d$ is not a prime number, and if the lowest common divisor of $d,\ k$ is 1, then the 
matrices $P_{0},\ P_{k}$ have been shown to be mutually unbiased. In the {\bf odd case}, 
 when $d$ is not a prime number, this appears very useful to find more than 3 MUB.
 
 \begin{proposition}
 Let d be an odd integer. If $E:=\left\{ k_{j}\right\}\subset \left\{0,1,..., d-1\right\}$
 is such that the lowest common divisor of $d, \ k_{j}-k_{j'}$ is 1 for all $k_{j},\ k_{j'}
 \in E$, then the set 
 $$\left\{ \1_{d}, P_{k_{j}}\right\}_{k_{j}\in E}$$ 
 defines a MUB.
 \end{proposition}
 
 Proof: The proof is quite simple and uses Theorem 2.11 (ii). Namely, since $P_{k}= D^{-k}P_{0}$,
 we have:
 \beq
 \label{biased}
 P_{k}^* P_{j}= P_{0}^* D^{k-j}P_{0}= P_{k-j}^*P_{0}
 \edq
 Now, this follows from Proposition 2.12.\\
 \sq\\

 \begin{corollary}
Let $d=mn,\ \mbox{with}\ n, m\in \mathbb N \ \mbox{prime numbers, and}\ n<m$.
Then the cardinality of the set of $d \times d$ unbiaised bases $N(d)$ satisfies:
$$N(d)\ge N(n)\equiv n+1$$ 
 \end{corollary}
 
 Proof: For n=2, we are in the {\bf even case} studied in the previous subsection. It has
 already been established that $N(d)\ge 3$ (Corollary 2.8). If $n$ is {\bf odd}, (then so is $m$),
 the matrices $P_{k}$ for $k \in F_{n}$ are all mutually unbiased. Thus we can choose as a MUM the set 
 $$\left\{\1_{d}, P_{0}, P_{1},..., P_{n-1}\right\}$$
 \sq\\
  
 \begin{remark}
 A similar, but apparently more general result, has been proven in \cite{klaro}.
 \end{remark}
 
 \bigskip
 \noindent
 {\bf EXAMPLE 1: d=15 :} There are 4 MUB's, defined either by 
 $$\left\{\1_{15}, P_{0}, P_{1}, P_{2}\right\}\ 
 \left\{\1_{15}, P_{0}, P_{2}, P_{4}\right\}\ 
 \left\{\1_{15}, P_{0}, P_{1}, P_{8}\right\}\ 
 \left\{\1_{15}, P_{0}, P_{4}, P_{8}\right\}\ 
 \left\{ \1_{15}, P_{0},  P_{7}, P_{14}\right\}$$
 
 \noindent
 {\bf EXAMPLE 2 : d=21} There are 4 MUB's, defined for example by
 $$\left\{\1_{21}, P_{0}, P_{1}, P_{2}\right\}$$
 Of course we do not know whether or not this is the maximum number of MUB's in these cases.

\subsection{THE PRIME NUMBER CASE}

\begin{proposition}
Let us assume that $d$ is a {\bf prime number}$\ge 3$. Then all unitary $d \times d$ matrices
$P_{0}^*P_{k}, \ k \in \left\{0,1,...,d-1\right\}$ are unbiased.
\end{proposition}

Proof: Any prime number $\ge 3$ being odd, the result is a consequence of Lemma 2.13, 
since then any $k \in F_{d}$ is relatively prime to $d$.\\
\sq\\

\begin{theorem}
for $d$ a {\bf prime number}, the following set of matrices
$$\left\{\1_{d}, D^{-k}P_{0},\ k=0,1,...,d-1\right\}$$
defines a maximal set of MUM.
\end{theorem}

Proof: We use Theorem 2.11: thus $P_{k}= D^{-k}U$, so that
$$P_{k}^*P_{j}= P_{0}^*D^{k-j}P_{0}= P_{k-j}^*P_{0}$$
so that if $j \ne k$ the result follows from Proposition 2.12.\\
\sq\\

\begin{remark}
The fact that in dimension $d$ there is at most $d+1$ MUB, and exactly $d+1$ for $d$ a
prime number is known for a long time. See for example \cite{wootters} and references
herein contained.
\end{remark}

\mysection{THE CASE WHERE d IS THE SQUARE OF A PRIME NUMBER}

Consider the Tensor-Product $d^2 \times d^2$ matrices introduced by Kibler-Planat \cite
{kipla}, (here restricted to two-tensor products):

\beq
\label{W}
W_{j,k}:= V_{j}^{(d)}\otimes V_{k}^{(d)},\ j,k \in \left\{0,1,...,d-1\right\}
\edq
where $d$ is a {\bf prime number greater than or equal to 3}, and $V_{j}^{(d)}$ is the
corresponding $d \times d$ matrices, for $j \in \left\{0,1,...,d-1\right\}$.\\
Let $U^{(d)}:= {\rm diag}(1,q,...,q^j,...,q^{d-1})$ where $q$ is defined by (\ref{q}), and
$U$ be the $d^2\times d^2$ diagonal unitary matrix
$$U:= U^{(d)}\otimes U^{(d)}$$
Consider the unitary matrices $P_{k}^{(d)}$ constructed in the previous section, and let
for $j,k \in \left\{0,1,...,d-1\right\}$ the $d^2\times d^2$ unitary matrices
$$P_{j,k}:= P_{j}^{(d)}\otimes P_{k}^{(d)}$$
Then we have:

\begin{proposition}
$$W_{j,k}P_{j,k}= P_{j,k}U$$
\end{proposition}

Proof: This immediately follows (omitting the superscript d for simplicity) from:
$$(V_{j}\otimes V_{k})\  (P_{j}\otimes P_{k})= (V_{j}P_{j})\ \otimes\ (V_{k}P_{k})
= (P_{j}U)\ \otimes (P_{k}U)\equiv P_{j,k}U$$
\sq\\

\begin{theorem}
(i) The matrices $P_{j,k}$ are unbiased $d^2\times d^2$ matrices, $\forall j,k \in
\left\{0,1,...,d-1\right\}$.\\
(ii) For any $j,k \in \left\{0,1,...,d-1\right\},\ k\ne j$, we have that
$P_{j,j},\ P_{k,k}$ are mutually  unbiased $d^2\times d^2$ matrices.
\end{theorem}

Proof: Recall that the ``tensor-product formalism'' enables to write the $d^2 \times d^2$
matrices as $2 \times 2$ block forms of $d \times d$ matrives. Namely $\forall j,k \in F_{d}$,
$$W_{j,k}\equiv \left(
\begin{array}{cc}
0& (-i)^j V_{k}\\
V_{k}&0
\end{array}\right)\quad P_{j,k}\equiv \frac{1}{\sqrt 2}\left(
\begin{array}{cc}
P_{k}&P_{k}\\
i^jP_{k}&-i^jP_{k}\end{array}\right)\quad U_{j,k}\equiv \left(
\begin{array}{cc}
(-i)^jU_{k}&0\\
0& -(-i)^jU_{k}\end{array}\right)$$
with $U_{k}$ diagonal matrices such that
$$V_{k}= P_{k}U_{k}P_{k}^*$$
Then the result follows from Proposition 2.20.\\
\sq\\

\begin{remark}
The above result provides only $d(d-1)/2$ MUB. But it is known (see \cite{wootters}, \cite{kipla})
that the maximun number which is here $d^2+1$ is attained. There is a ``trick'', not explained
here which allows to construct the ``missing'' bases, not only for the {\bf square of prime
numbers}, but more generally for {\bf any power of prime numbers.} We shall give the
explicit construction for $d=4$ in Chapter 5.
\end{remark}

\mysection{DIMENSIONS 2 AND 3}

$$P_{0}= \frac{1}{\sqrt 2}\left(
\begin{array}{cc}
1&1\\
1&-1
\end{array}\right)\qquad P_{1}= \frac{1}{\sqrt 2}\left(
\begin{array}{cc}
1&1\\
i&-i
\end{array}\right)$$

\begin{proposition}
(i) The sets $$E_{2}:= \left\{\1_{2}, P_{0}, P_{1}\right\},
\qquad E'_{2}:= \left\{\1_{2}, P_{1}, P_{1}^*\right\}$$
are complete MUM in dimension d=2.\\
(ii) The bases in $\mathbb C^2$ defined by $E_{2}\ \mbox{and}\ E'_{2}$ are the same 
MUB in dimension d=2.
\end{proposition}

Proof: (i) results from Propositions 2.5 and 2.7, for $E_{2}$, and for $E'_{2}$ from the fact that
$P_{1}^2$ is unbiased (in other words $P_{1}$ is mutually unbiased to itself). Namely:
$$P_{1}^3 =e^{-i\pi/4}\1_{2} $$
which implies that $P_{1}^2= e^{-i\pi/4}P_{1}^*$ which is unbiased.\\
(ii) Denote by $e_{1}:= \left(
\begin{array}{c}1\\
0\end{array}\right)$ and $e_{2}:= \left(
\begin{array}{c}0\\
1\end{array}\right)$ the natural basis in $\mathbb C^2$. Then the MUB defined by $E_{2}
,\quad E'_{2}$ are $\left\{B_{0}, B_{1}, B_{2}\right\}$ where
$$B_{0}:= \left\{e_{1}, e_{2}\right\}\quad B_{1}:=\left\{\frac{1}{\sqrt 2}
(e_{1}\pm e_{2})\right\}\quad B_{2}:= \left\{\frac{1}{\sqrt 2}(e_{1}\pm ie_{2})
\right\}$$
\sq\\

For the case of dimension $d=3$ we simply use Theorem 2.8 (ii) for the simple construction
of $P_{j},\ j\in \left\{0,1,2\right\}$:\\
 Let $q= \exp(\frac{2i\pi}{3})$
 $$P_{0}= \frac{1}{\sqrt 3}\left(
 \begin{array}{ccc}
 1&1&1\\
 1&q&q^2\\
 1&q^2&q
 \end{array}\right)\quad
 P_{1}= \frac{1}{\sqrt 3}\left(
 \begin{array}{ccc}
 1&1&1\\
 q^2&1&q\\
 1&q^2&q
 \end{array}\right)\quad
 P_{2}= \frac{1}{\sqrt 3}\left(
 \begin{array}{ccc}
 1&1&1\\
 q&q^2&1\\
 1&q^2&q
 \end{array}\right)$$
 
 \begin{proposition}
 (i) The set $E_{3}:= \left\{\1_{3}, P_{0}, P_{1}, P_{2}\right\}$ defines a maximal
 MUM for d=3.
 \\
 (ii) Define:
$$ P'_{1}:= \frac{1}{\sqrt 3}\left(
 \begin{array}{ccc}
 1&1&q\\
1&q&1\\
q&1&1
\end{array}\right )$$
Then the set $E'_{3}:= \left\{\1_{3}, P_{0}, P_{1}', P_{1}^{'*}\right\}$ defines a maximal
MUM in dimension d=3.
 \end{proposition}
 
 Proof: (i) simply follows from Theorem 2.14. Furthermore $E'_{3}$ defines the same
 MUB as $E_{3}$, which establishes (ii).\\
 \sq\\
 
 \mysection{THE CASE OF DIMENSION 4}
 
 There is nothing new in the results of this section (see \cite{band}, \cite{kipla},
  \cite{wootters}). The only point is that we construct explicit matrices that allow to
  complete the set of MUM provided in Section 3.\\
  According to Theorem 3.2, we have that $P_{0,0},\ P_{1,1}$ are mutually unbiased matrices.
  \\
  However $P_{0,1}, \ P_{1,0}$ are not mutually unbiased, neither to each other, nor to the
  two previous ones. TThe trick is to consider that the eigenspaces of $W_{0,1},\ W_{1,0}$
  with eigenvalues $\pm i$ are degenerate, so that vectors of these eigenspaces can be recombined
  to build MUB's.\\
  Namely take
  $$P'_{0,1}:= \frac{1}{\sqrt 2}\left(
  \begin{array}{cc}
  P_{0}&P_{0}\\
  -iP'_{0}&iP'_{0}
  \end{array}\right)\quad P'_{1,0}:= \frac{1}{\sqrt 2}\left(
  \begin{array}{cc}
  P_{1}&P_{1}\\
  -P'_{1}&P'_{1}
  \end{array}\right)\quad P'_{0,0}\equiv P_{0,0}\quad P'_{1,1}\equiv P_{1,1}$$
  with
  $$P'_{0}:= \frac{1}{\sqrt 2}\left(
  \begin{array}{cc}
  1&1\\
  -1&1
  \end{array}\right)\quad P'_{1}:= \frac{1}{\sqrt 2}\left(
  \begin{array}{cc}
  1&1\\
  -i&i
  \end{array}\right)$$
  Actually, defining the unitary $4\times 4$ matrix (that commutes with $U_{1,0}\ 
  \mbox{and}\ U_{0,1}$) as
  $$A:= \frac{e^{-i\pi/4}}{\sqrt 2}\left(
  \begin{array}{cccc}
  1&0&0&i\\
  0&1&i&0\\
  0&i&1&0\\
  i&0&0&1
  \end{array}\right)$$
  we have:
  $$P_{1,0}= P'_{1,0}A\qquad P_{0,1}= P'_{0,1}A^*$$
  Then
  
  \begin{proposition}
  $$W_{0,1}P'_{0,1}= P'_{0,1}U_{0,1}, \quad W_{1,0}P'_{1,0}= P'_{1,0}U_{1,0}$$
  and $P_{i,j}^{'*}P'_{k,l}$ are unbiased matrices $\forall (i,j)\ne (k,l)\ i,j,k,l \in \left\{
  0,1\right\}$.
  \end{proposition}
  
  Proof: We check that $P_{0,1}^{'*}P_{1,0}$ is an unbiased matrix. We have:
  $$P_{0,1}^{'*}P_{1,0}=\frac{1}{2}\left(
  \begin{array}{cc}
 ( P_{0}^*-iP_{0}^{'*})P_{1}&(P_{0}^*+iP_{0}^{'*})P_{1}\\
(P_{0}^*+iP_{0}^{'*})P_{1}& (P_{0}^* -iP_{0}^{'*})P_{1}
\end{array}\right )$$
But
$$(P_{0}^*-iP_{0}^{'*})P_{1}= \left(
\begin{array}{cc}1&-i\\
-i&1
\end{array}\right)\qquad (P_{0}^*+iP_{0}^{'*})P_{1}= \left(
\begin{array}{cc}
i&1\\
1&i
\end{array}\right)$$
The other cases can be shown similarly.\\
\sq\\

 \mysection{THE CASE OF DIMENSION 6}
 It is le least even dimension which in not {\bf the power of a prime number}. 
 Let $j := \exp(\frac{2i\pi}{6})$. Then
 $$P_{0}= \frac{1}{\sqrt 6}\left(
 \begin{array}{cccccc}
 1&1&1&1&1&1\\
 1&j&j^2&-1&-j&-j^2\\
 1&j^2&-j&1&j^2&-j\\
 1&-1&1&-1&1&-1\\
 1&-j&j^2&1&-j&j^2\\
 1&-j^2&-j&-1&j^2&j
 \end{array}\right)
 \quad P_{1}= \frac{1}{\sqrt 6}\left(
 \begin{array}{cccccc}
 1&1&1&1&1&1\\
 -ij^2&i&ij&ij^2&-i&-ij\\
 1&j^2&-j&1&j^2&-j\\
 -i&i&-i&i&-i&i\\
 j^2&1&-j&j^2&1&-j\\
 -i&ij^2&ij&i&-ij^2&-ij
 \end{array}\right)$$
 
 \begin{lemma}
 Let $\tilde D$ be the following unitary diagonal matrix:
 $$\tilde D := {\rm diag}(1, -ij^2, 1, -i, j^2, -i)$$
 Then we have:
 $$P_{1}= \tilde D P_{0}$$
 \end{lemma}
 
 \begin{proposition}
 The set $E_{6}:= \left\{\1_{6}, P_{0}, P_{1}\right\}$ defines a  MUM in dimension
 d=6.
 \end{proposition}
 
 Proof: This follows simply from Ptoposition 2.5 and  Proposition 2.7. Moreover we have:
 $$P_{0}^*VP_{0}= U\qquad P_{1}^* V_{1}P_{1}= iU$$
 \sq\\
 
 \begin{remark}
 The fact that $N(6)=3$ is the maximum number of MUB in dimension 6 is a conjecture 
 apparently due to Zauner \cite{zau}. Some progress has been recently made in dimension 6 
 by M. Grassl \cite{gra}.
 \end{remark}

 \mysection{THE CASE OF DIMENSIONS 12 AND 20}
 
 Let $d= 4m$ where $m$ is an odd number $\ge 3$. Then consider the $4 \times 4$ matrices
  $W_{k}, \ k = 0,1,...,3$ constructed in Section 3, together with the set of matrices $V_{k},
   \ k\in F_{m}$
  constructed in Subsection 2.3. Denote by $Q_{j},\ j=0,1,...,3$ the unitary $ 4 \times 4$
  matrices 
  $P_{k,l},\ 
  k,l \in \left\{0,1\right\}$, (in lexicographic order) provided in Section 5 for $d=4$,
   and by $P_{j}, \ j\in F_{m}$ the $m \times m$ unitary matrices constructed
  in Subsection 2.3. Then one has:
  
  \begin{lemma}
  For any $j= 0,1,..., {\rm Inf}(4, m+1)$, there exists a {\bf diagonal matrix} $U_{j}$ such that
  $$(W_{j}\otimes V_{j})\ (Q_{j}\otimes P_{j})= (Q_{j}\otimes P_{j})U_{j} $$
  \end{lemma}
  
  The proof is very similar to the one provided in Section 3. Furthemore the idea of tensor-product
  methods in this situation is already present in \cite{klaro}.\\
  
  \noindent
  Actually the new ingredient in this Section is to establish  explicit $4m\times 4m$ matrices
   $R_{j}:= Q_{j}\otimes P_{j} $  in $4 times 4$ or $m \times m$
  block forms; let us specify them for $m=3,\ m=5$:
  \begin{lemma}
 (i) Let $d=12$. Thus $m=3$ and denoting by $q$ the 3rd root of unity $q:=\exp(2i\pi/3)$,
 we have:
 $$R_{0}:= \frac{1}{\sqrt 3}\left(
 \begin{array}{ccc}
 Q_{0}&Q_{0}&Q_{0}\\
 Q_{0}&qQ_{0}&q^2Q_{0}\\
 Q_{0}&q^2Q_{0}&qQ_{0}
 \end{array}\right)\quad R_{1}:= \frac{1}{\sqrt 3}\left(
 \begin{array}{ccc}
 Q_{1}&Q_{1}&Q_{1}\\
 q^2Q_{1}& Q_{1}&qQ_{1}\\
 Q_{1}&q^2Q_{1}&qQ_{1}
 \end{array}\right)$$
 $$R_{2}:= \frac{1}{\sqrt 3}\left(
 \begin{array}{ccc}
 Q_{2}&Q_{2}&Q_{2}\\
 qQ_{2}&q^2Q_{2}&Q_{2}\\
 Q_{2}&q^2Q_{2}&qQ_{2}
 \end{array}\right)$$
 The matrices $R_{j},\ j=0,1,2$ are obviously unbiased unitary matrices and are mutually
 unbiased. Thus the set $\left\{\1_{12}, R_{0}, R_{1}, R_{2}\right\}$ defines a
 set of 4 MUB's for $d=12$. Futhermore any choice of $Q_{j}$'s among the 4 matrices
 $P_{j,k}, j,k \in \left\{0,1\right\}$ (not necessarily the lexicographic order)
 gives the same result, but not the same MUB's.\\
 
 (ii) Let $d=20$, thus $m=5$. Take 4 unitary $5 \times 5$ matrices among the 6 possible
  $P_{j}$'s
 in dimension 5. Then we have:
 $$R'_{0}:= \frac{1}{2}\left(
 \begin{array}{cccc}
P_{0}& P_{0}&P_{0}&P_{0}\\
P_{0}&-P_{0}&P_{0}&-P_{0}\\
P_{0}&P_{0}&-P_{0}&-P_{0}\\
P_{0}&-P_{0}&-P_{0}&P_{0}\end{array}\right )\quad R'_{1}:=\frac{1}{2}\left(
\begin{array}{cccc}
P_{1}&P_{1}&P_{1}&P_{1}\\
P_{1}&-P_{1}&P_{1}&-P_{1}\\
-iP_{1}&-iP_{1}&iP_{1}&iP_{1}\\
iP_{1}&-iP_{1}&-iP_{1}&iP_{1}
\end{array}\right)$$
$$R'_{2}:= \frac{1}{2}\left(
\begin{array}{cccc}
P_{2}&P_{2}&P_{2}&P_{2}\\
iP_{2}&-iP_{2}&iP_{2}&-iP_{2}\\
-P_{2}&-P_{2}&P_{2}&P_{2}\\
iP_{2}&-iP_{2}&-iP_{2}&iP_{2}\\
\end{array}\right)\quad R'_{3}:= \frac{1}{2}\left(
\begin{array}{cccc}
P_{3}&P_{3}&P_{3}&P_{3}\\
iP_{3}&-iP_{3}&iP_{3}&-iP_{3}\\
iP_{3}&iP_{3}&-iP_{3}&-iP_{3}\\
-P_{3}&P_{3}&P_{3}&-P_{3}
\end{array}\right)$$
Then the $20 \times 20$ unitary matrices $R'_{j},\ j=0,1,...3$ are
 unbiaised and mutually unbiased. Thus the set
 $$\left\{\1_{20}, R'_{0}, R'_{1}, R'_{2}, R'_{3}\right\}$$
 defines a set of 5 MUB's. 
 \end{lemma}

{\bf Acknowledgements :} It is a pleasure to thank M. Kibler for learning me everything
about MUB's, providing me with \cite{kipla} before publication and for his careful reading
of this manuscript. I am also indebted
to F. Moulin and J. Marklof for useful informations and comments about Gauss Sums.

\end{document}